\documentclass[showpacs,amsmath,amssymb,aps,showkeys,floatfix,prd,a4paper]{revtex4}

\usepackage[dvips]{graphicx}
\usepackage{dcolumn}
\usepackage{bm}
\usepackage{epsfig}
\usepackage{amsfonts}
\usepackage{amssymb,amscd}

\def\lsim{\raise0.3ex\hbox{$<$\kern-0.75em\raise-1.1ex\hbox{$\sim$}}}

\def\gsim{\raise0.3ex\hbox{$>$\kern-0.75em\raise-1.1ex\hbox{$\sim$}}}

\newcommand{\be}{\begin{equation}}

\newcommand{\ee}{\end{equation}}

\def\beq{\begin{equation}}

\def\eeq{\end{equation}}

\def\beqa{\begin{eqnarray}}

\def\eeqa{\end{eqnarray}}

\newcommand{\ba}{\begin{eqnarray}}

\newcommand{\ea}{\end{eqnarray}}

\def\gappeq{\mathrel{\rlap {\raise.5ex\hbox{$>$}}

{\lower.5ex\hbox{$\sim$}}}}

\def\lappeq{\mathrel{\rlap{\raise.5ex\hbox{$<$}}

{\lower.5ex\hbox{$\sim$}}}}

\def\Toprel#1\over#2{\mathrel{\mathop{#2}\limits^{#1}}}

\begin{document}

\title{Photoproduction of relativistic QED bound states in hadronic collisions}
\author{R. Francener}
\affiliation{Departamento de F\'isica, Universidade do Estado de Santa Catarina, 89219-710 Joinville, SC, Brazil. }
\author{V.P. Gon\c{c}alves}
\affiliation{High and Medium Energy Group, Instituto de F\'{\i}sica e Matem\'atica,  Universidade Federal de Pelotas (UFPel)\\
Caixa Postal 354,  96010-900, Pelotas, RS, Brazil.}
\author{B.D.  Moreira}
\affiliation{Departamento de F\'isica, Universidade do Estado de Santa Catarina, 89219-710 Joinville, SC, Brazil. 
}

\begin{abstract}
In this paper we investigate the photoproduction of  QED bound states in proton -- proton, proton -- nucleus and nucleus -- nucleus collisions at RHIC, LHC and FCC energies   considering an accurate treatment of the absorptive corrections and for the nuclear form factor. The total cross sections for the production of singlet and triplet $(l^+ l^-)$  states, with $l = e,\, \mu, \, \tau$, are estimated and a detailed analysis of the photoproduction of singlet QED bound states is performed  considering the rapidity ranges covered by central and forward detectors. The impact of the Coulomb corrections on the parapositronium production in heavy ion collisions is estimated. We predict a large number of events associated to the production of  ($e^+ e^-$) and ($\mu^+ \mu^-$) states in hadronic collisions. 
\end{abstract}

\pacs{12.38.-t, 24.85.+p, 25.30.-c}

\keywords{Quantum Electrodynamics, QED Bound States, Ultraperipheral Collisioms.}

\maketitle

\vspace{1cm}

\section{Introduction}
The LHC experiments have focused part of its physics goals into the particle production by photon - induced interactions in order to improve our understanding of the Standard Model (SM) and search for signals of New Physics \cite{review_forward,Bruce:2018yzs}. Such studies are strongly motivated by the large photon - photon luminosity ($\propto Z_1^2 Z_2^2$, where $Z_1$ and $Z_2$ are the atomic numbers of the incident particles) and the clean environment present in ultraperipheral hadronic collisions \cite{upc}, which enhance the production cross sections and reduce the  backgrounds, allowing to derive stringent limits on some Beyond Standard Model (BSM) scenarios that predict  new particles that couple predominantly to photons. Another route to searching for a signal of New Physics is to study with high precision the interactions of the Standard Model, by looking for small deviations in the SM predictions. Recent studies have pointed out that the study of bound states of leptons provide a probe that is sensitive to beyond SM physics \cite{true_lhcb,true_fixed} and is  an ideal testing ground of QED, since it allows to test the properties of leptons,  the charge -- conjugation, parity and time -- reversal (CPT) invariance of the theory as well allows to study the bound state physics (See e.g. Refs. 
  \cite{Stroscio:1975fa, Brodsky:2009gx, Lamm:2013oga, Wiecki:2014ola,Hoyer:2016aew, Bass:2019ibo,Mondal:2019rhs}). Such results motivate the analysis of the photoproduction of QED bound states in hadronic collisions.

\begin{figure}[t]
	\centering
	\begin{tabular}{ccc}
	\includegraphics[width=0.45\textwidth]{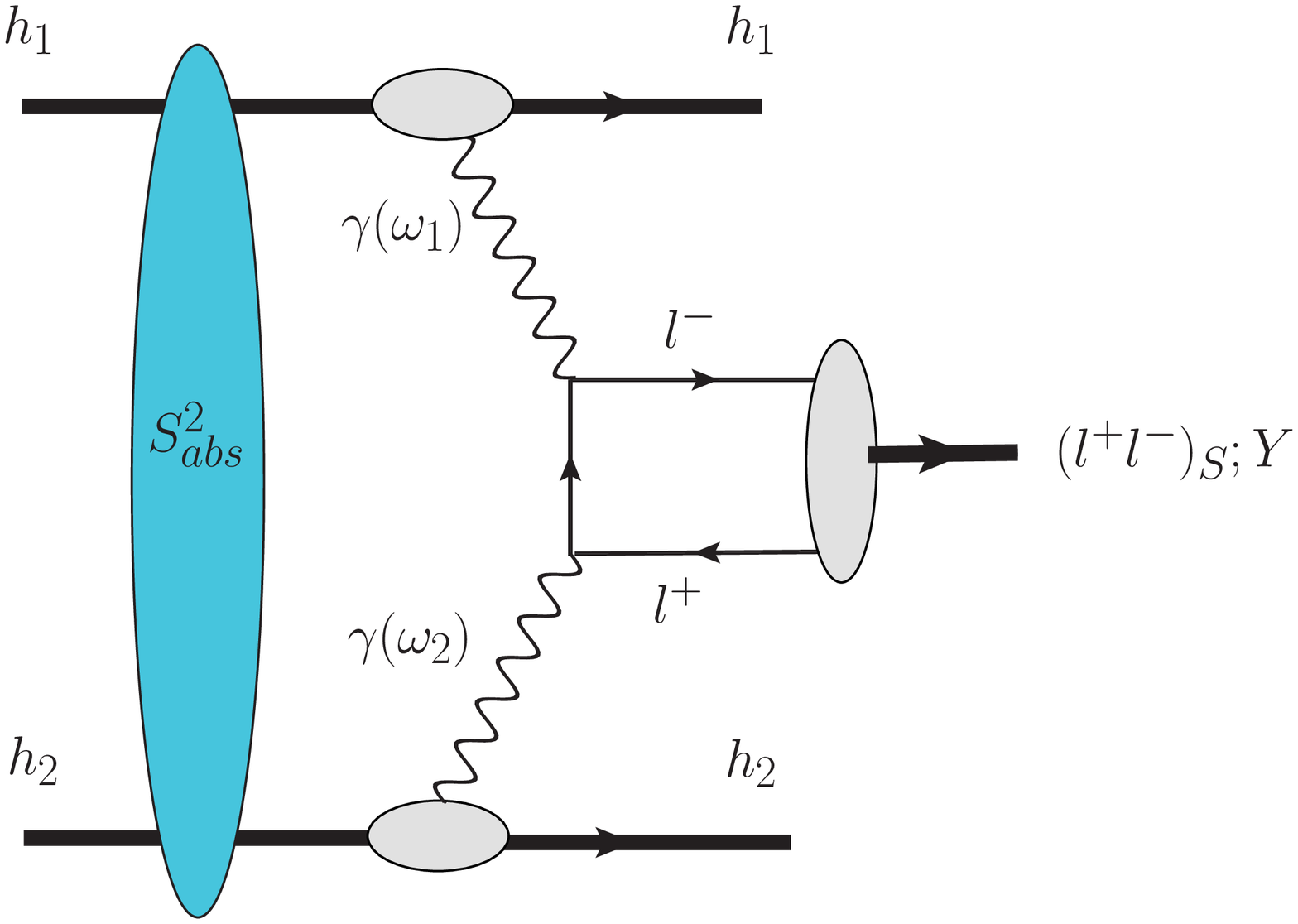} & \,\,\,\,\, & \includegraphics[width=0.45\textwidth]{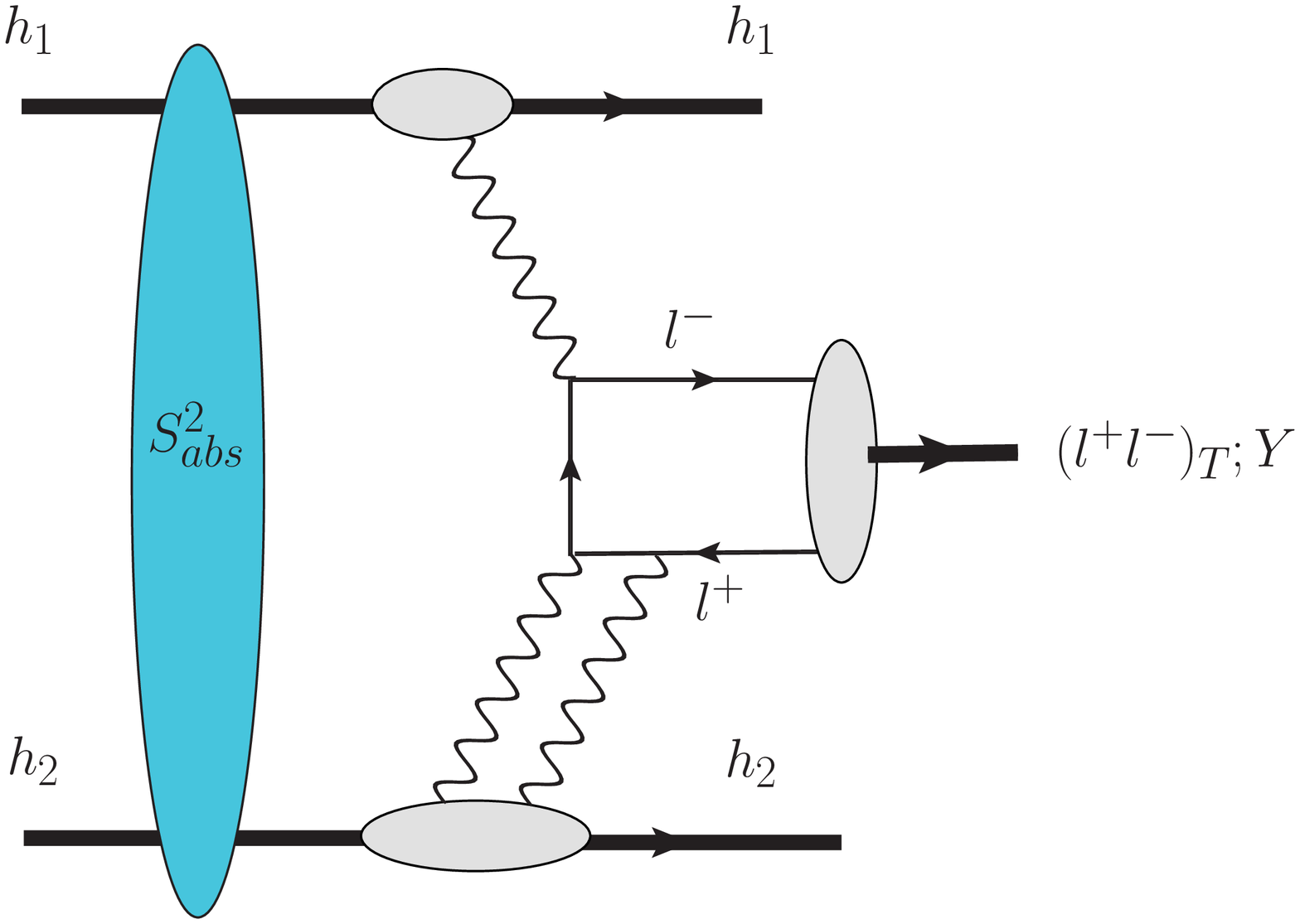} \\
	(a) & \, &  (b) 
			\end{tabular}
\caption{Photoproduction of (a) singlet and (b) triplet QED bound states  at Born level in hadronic collisions.}
\label{fig:diagram_Born}
\end{figure}

The structure of bound states of leptons is very similar to that of hydrogen. In particular, its ground states can be in a singlet state with spins antipallel and total spin $s=0$, or it can be in the triplet state with spins parallel and total spin $s=1$. Such singlet and triplet states are usually denoted para and ortho QED bound states, respectively. An important difference from hydrogen is  that for a bound state formed by leptons of a same flavour, denoted hereafter by $(l^+l^-)$, annihilation can occur, with the number of photons $n$ emitted in the decay process being governed by the charge -- conjugation selection rule $(-1)^{l + s} = (-1)^n$, where $l$ is the orbital angular momentum. Therefore, for a QED bound state in the ground state, a singlet state decays into an even number of photons, while the triplet must decay into an odd number. In addition, one has that these states can be produced at the Born level by two  and three photon fusion. As a consequence, such states can be  studied in hadronic collisions, since the incident charged hadrons act as  sources of almost real photons and photon - induced interactions may happen \cite{upc}.  The typical diagrams for the photoproduction of QED bound states in hadronic collisions are represented in Fig. \ref{fig:diagram_Born}. In the equivalent photon approximation \cite{epa}, the associated total cross sections for the photoproduction of singlet (S) and triplet (T) QED bound states can be expressed, respectively, by
\begin{eqnarray}
\sigma \left[h_1 h_2 \rightarrow h_1  (l^+ l^-)_S  h_2;s_{NN} \right]   
&=& \int \mbox{d}^{2} {\mathbf r_{1}}
\mbox{d}^{2} {\mathbf r_{2}} 
\mbox{d}\omega_1 
\mbox{d} \omega_2 \,   N\left(\omega_{1},{\mathbf r_{1}}  \right )
 N\left(\omega_{2},{\mathbf r_{2}}  \right ) \, \hat{\sigma}\left[\gamma \gamma \rightarrow (l^+ l^-)_S ; 
W_{\gamma \gamma} \right] S^2_{abs}({\mathbf b})  
  \,\,\, ,
\label{Eq:cs_singlet}
\end{eqnarray}
and
\begin{eqnarray}
\sigma \left[h_1 h_2 \rightarrow h_1  (l^+ l^-)_T  h_2;s_{NN} \right]   
&=& \int \mbox{d}^{2} {\mathbf r_{1}}
\mbox{d}\omega_1 
 \,  N_1\left(\omega_{1},{\mathbf r_{1}}  \right )\, \hat{\sigma}\left[\gamma h_2 \rightarrow (l^+ l^-)_T h_2 ; 
W_{\gamma h_2} \right]  S^2_{abs}({\mathbf b})  + (1 \leftrightarrow   2) 
  \,\,\, ,
\label{Eq:cs_triplet}
\end{eqnarray}
where $\sqrt{s_{NN}}$ is center - of - mass energy of the $h_1 h_2$ collision, $N(\omega_i, {\mathbf r}_i)$ is the equivalent photon spectrum  
of photons with energy $\omega_i$ at a transverse distance ${\mathbf r}_i$  from the center of hadron, defined in the plane transverse to the trajectory, and  $\hat{\sigma}$ represents the cross section for the photon - induced interaction for a  given photon - photon (photon - hadron) center - of - mass energy $W_{\gamma \gamma}$ ($W_{\gamma h}$). Moreover, the factor $S^2_{abs}({\mathbf b})$ depends on the impact parameter ${\mathbf b}$ of the hadronic collision and  is denoted the absorptive  factor, which excludes the overlap between the colliding hadrons and allows to take into account only ultraperipheral collisions (UPCs), characterized by $|{\mathbf b}| \ge R_{h_1} + R_{h_2}$, where $R_{h_i}$ is the hadron radius. These collisions are dominated by photon - induced interactions and the photon -- photon luminosity  is high, increasing with $Z_1^2 Z_2^2$. Such property has motivated the studies performed in Refs. \cite{serbo_jtep, serbo_pra,serbo_muonium,serbo_posi,nos_muonium}, which have analyzed the photoproduction of the positronium ($e^+ e^-$) and muonium ($\mu^+ \mu^-$) states in heavy ion collisions. One of our goals is to  extend these previous studies for a ($\tau^+ \tau^-$) bound state, the tauonium. We also will update the predictions for the photoproduction of singlet and triplet QED bound states in heavy ion collisions at RHIC and LHC and present, for the first time, results for the energies of the High -- Energy LHC \cite{he_lhc} and Future Circular Collider (FCC)  \cite{fcc}. Moreover, we will perform a detailed analysis of the  photoproduction of singlet QED bound states in pp, pA and AA collisions and the cross sections will be estimated considering the rapidity ranges usually covered by central ($-2.5 \le Y \le 2.5$) and forward ($2 \le Y \le 4.5$) detectors. Finally, the impact of the Coulomb corrections \cite{serbo_posi,serbo_pra}, associated to multiphoton exchange, on the predictions for the parapositronium production in heavy ion collisions will also be discussed. { It is important to emphasize that recent results, obtained in Ref. \cite{Zha:2021jhf}, indicate that higher order corrections are needed to  describe the QED pair production in UPCs}.
  Our study  is strongly motivated by the fact that the resulting  final state is very clean, consisting  of a QED bound state,  two intact hadrons and  two rapidity gaps, i.e. empty regions  in pseudo-rapidity that separate the intact very forward nuclei from the $(l^+ l^-)$ state. Such aspects can, in principle, be used to separate the events and to probe the QED bound states.

\section{Photoproduction of QED bound states}
The basic ingredients to estimate the total cross sections for the photoproduction of singlet and triplet QED bound states in hadronic collisions, expressed by the Eqs. (\ref{Eq:cs_singlet}) and (\ref{Eq:cs_triplet}), are the equivalent photon spectrum $N(\omega_i, {\mathbf r}_i)$, the absorptive factor $S^2_{abs}({\mathbf b})$ and the photon - photon ($\hat{\sigma}_{\gamma \gamma}$) and photon - hadron ($\hat{\sigma}_{\gamma h}$) cross sections. In what follows, we will discuss each one of these elements. One has that the
photon spectrum can be 
expressed in terms of the charge form factor $F(q)$ as follows \cite{upc}
\begin{eqnarray}
 N(\omega_i,r_i) = \frac{Z^{2}\alpha}{\pi^2}\frac{1}{r_i^{2} v^{2}\omega_i}
\cdot \left[
\int u^{2} J_{1}(u) 
F\left(
 \sqrt{\frac{\left( \frac{r_i\omega_i}{\gamma_L}\right)^{2} + u^{2}}{r_i^{2}}}
 \right )
\frac{1}{\left(\frac{r_i\omega_i}{\gamma_L}\right)^{2} + u^{2}} \mbox{d}u
\right]^{2} \,\,,
\label{fluxo}
\end{eqnarray}
where $\alpha$ is the electromagnetic coupling constant, $\gamma_L$ is the Lorentz factor and $v$ is the hadron velocity.
For a nucleus, we will estimate the associated photon spectrum considering the realistic form factor, which corresponds to the Wood - Saxon distribution and is the Fourier transform of the charge density of the nucleus, constrained by the experimental data. It can be analytically expressed by
\begin{eqnarray}
 F(q^{2}) = 
 \frac{4\pi\rho_{0}}{Aq^{3}} 
 \left[ 
 \sin(qR) - qR \cos(qR) 
 \right]
 \left[
 \frac{1}{1 + q^{2} a^{2}}
 \right]
\end{eqnarray}
with $a = 0.549 \,(0.535)$ fm and $R_{A} = 6.63 \, (6.38)$ fm for A = Pb (Au)  \cite{DeJager:1974liz,Bertulani:2001zk}. On the other hand, for a proton, we will assume that 
\begin{eqnarray}
	F(q^{2})=\dfrac{\Lambda^{4}}{(\Lambda^{2}+q^{2})^{2}} \,\,,
	\label{eq4}
\end{eqnarray}
with $\Lambda^{2} = 0.71 \mathrm{GeV}^{2}$ \cite{dz}. { As demonstrated, e.g.  in Ref. \cite{Zha:2021jhf}, the lowest order QED predictions derived using the  standard dipole form factor of proton are able to describe the current data for the exclusive pair production in $pp$ collisions}.
For the absorptive factor $S^2_{abs}({\mathbf b})$ we will assume that it is given by   \cite{Baur_Ferreira}
\begin{eqnarray}
S^2_{abs}({\mathbf b})=\Theta (|{\mathbf b}| - R_{h_1} - R_{h_2}).
	\label{eq5}
\end{eqnarray}
Such factor  suppress the contributions associated to strong interactions by excluding the overlap between the colliding hadrons. As discussed in detail in Ref. \cite{nos_dilepton}, the absorptive factor can also be estimated using the Glauber approach (See e.g. Ref. \cite{Baltz_Klein}), which is more realistic for  $|{\mathbf b}| \approx R_{h_1} + R_{h_2} $. However, for the production of systems with small invariant masses, which is the case of QED bound states, the difference between the different models for 
$S^2_{abs}({\mathbf b})$ is negligible \cite{nos_dilepton}. At the Born level, the cross section $\hat{\sigma}_{\gamma \gamma}$ for the photoproduction of a singlet QED bound state in $\gamma \gamma$ interactions can be estimated using the Low formula \cite{Low}, which allow us to express this cross section in terms of the two-photon decay width $\Gamma_{(l^+ l^-)_S \rightarrow \gamma \gamma}$ as  follows
\begin{eqnarray}
 \hat{\sigma}_{\gamma \gamma \rightarrow (l^+ l^-)_S}(\omega_{1},\omega_{2}) = 
8\pi^{2} (2J+1) \frac{\Gamma_{(l^+ l^-)_S \rightarrow \gamma \gamma}}{M} 
\delta(4\omega_{1}\omega_{2} - M^{2}) \, ,
\label{Low_cs}
\end{eqnarray}
where $M = 2 m_{l}$ and $J$ are, respectively, the mass and spin of the  produced singlet QED state. In the non -- relativistic approximation, one have that only the probability density of $s$ -- states at the origin does not vanish, which implies that $|\Psi_{ns}(0)|^2 = \alpha^3 m_{l}^3/8 \pi n^3$. Consequently, one has $\Gamma (n \,^1S_0) = \alpha^5 m_{l}/2 n^3$ and that the $\gamma \gamma$ cross section for the lowest singlet QED bound state is given by
\begin{eqnarray}
 \hat{\sigma}_{\gamma \gamma \rightarrow (l^+ l^-)_S}(\omega_{1},\omega_{2}) = 2 \pi^2 \alpha^5 \delta(4\omega_{1}\omega_{2} - M^{2})\,\,.
\end{eqnarray}
Finally, the cross section $\hat{\sigma}_{\gamma h}$ for the photoproduction of a triplet QED bound state in $\gamma h$ interactions was estimated in Ref. \cite{serbo_pra}, being given by
\begin{eqnarray}
\hat{\sigma}_{\gamma h \rightarrow (l^+ l^-)_T h} (W_{\gamma h}^2 = 2 \omega \sqrt{s_{NN}})  = 4 \nu^2 \sigma_{0} \zeta (3) B(\nu)\,\,,
	\label{eq7}
\end{eqnarray}
where $\nu = Z \alpha$, $\sigma_0 = \pi \nu^2 \alpha^4/m_l^2$ and the function $B(\nu)$ is 
\begin{eqnarray}
B(\nu) = \left(\frac{\pi \nu}{\sinh \pi \nu}\right)^2 \frac{1}{8} \int_0^1 dz\, \sqrt{z} (1 + z) [F(1+i\nu, 1-i\nu;2;z]^2 \,\,,
\end{eqnarray}
with $F$ being the Gauss hypergeometric function.

\begin{center}
	\begin{table}[t]
		\begin{tabular}{|c|c|c|c|c|c|c|}
			\hline
			\hline 
			& {\bf Parapositronium}          &       {\bf Paramuonium} &    {\bf Paratauonium}   \\
			\hline 	
			\hline 
AuAu ($\sqrt{s_{NN}} = 0.2$ TeV)  & 112.1$\times 10^{12}$ (11.2$\times 10^{8}$) & 150.0$\times 10^{6}$ (1.5$\times 10^{3}$) & 3.8$\times 10^{3}$ (0.04) \tabularnewline
			\hline 
PbPb ($\sqrt{s_{NN}} = 5.02$ TeV)& 333.2$\times 10^{12}$ (33.3$\times 10^{8}$) & 1297.0$\times 10^{6}$ (12.9$\times 10^{3}$) & 832.7$\times 10^{3}$   (8.3) \tabularnewline
			\hline 
PbPb ($\sqrt{s_{NN}} = 10.6$ TeV) & 400.3$\times 10^{12}$ (40.0$\times 10^{8}$) & 1796.0$\times 10^{6}$ (17.9$\times 10^{3}$) & 1450.0 $\times 10^{3}$ (14.5) \tabularnewline
\hline
PbPb ($\sqrt{s_{NN}} = 39.0$ TeV) & 537.6$\times 10^{12}$ (59.1$\times 10^{9}$) & 2945.0$\times 10^{6}$ (32.4$\times 10^{4}$) & 3142.0$\times 10^{3}$ (345.6)    \tabularnewline
			\hline 
			\hline 
			\hline 
			& {\bf Orthopositronium}          &       {\bf Orthomuonium} &    {\bf Orthotauonium}   \\
			\hline 	
			\hline 
AuAu ($\sqrt{s_{NN}} = 0.2$ TeV)  & 6.7$\times 10^{12}$ (0.67$\times 10^{8}$) & 21.2$\times 10^{6}$ (0.22$\times 10^{3}$) & 0.02$\times 10^{3}$    (0.19$\times 10^{-3}$) \tabularnewline
			\hline 
PbPb ($\sqrt{s_{NN}} = 5.02$ TeV)& 23.3$\times 10^{12}$ (2.3$\times 10^{8}$) & 90.8$\times 10^{6}$  (0.91$\times 10^{3}$) & 0.40$\times 10^{3}$ (4.1$\times 10^{-3}$)       \tabularnewline
			\hline 
PbPb ($\sqrt{s_{NN}} = 10.6$ TeV) & 27.9$\times 10^{12}$ (2.8$\times 10^{8}$) & 110.5$\times 10^{6}$ (1.1$\times 10^{3}$) & 0.56$\times 10^{3}$  (5.6$\times 10^{-3}$)        \tabularnewline
\hline
PbPb ($\sqrt{s_{NN}} = 39.0$ TeV) & 37.0$\times 10^{12}$ (4.0$\times 10^{8}$) & 150.9$\times 10^{6}$ (1.6$\times 10^{3}$) & 0.89$\times 10^{3}$ (98.3$\times 10^{-3}$)        \tabularnewline
			\hline 
			\hline 			
		\end{tabular}
		\caption{Total cross sections in fb (Event rates per year)  for the production of singlet and triplet QED bound states in pp, pA and AA collisions.}
		\label{table:AA}
	\end{table}
\end{center}

In Table \ref{table:AA} we present our results for the total cross sections for the photoproduction of singlet and triplet QED bound states in heavy ion collisions derived at the Born level and considering the energies of the RHIC ($\sqrt{s_{NN}} = 0.2$ TeV), LHC ($\sqrt{s_{NN}} = 5.02$ TeV), HE -- LHC ($\sqrt{s_{NN}} = 10.6$ TeV) and FCC ($\sqrt{s_{NN}} = 39.0$ TeV). Moreover, we present our predictions for the number of events per year (in parenthesis) obtained assuming
the expected integrated luminosity  per year for the RHIC/LHC/HE-LHC   as being  10 nb$^{-1}$, and   110 nb$^{-1}$ for the FCC. As expected from Refs.  \cite{serbo_pra,serbo_muonium,serbo_posi} the singlet cross sections are larger than the triplet one, with the ratio $\sigma_T/\sigma_S$ being smaller for larger energies. We predict slightly larger values for the total cross sections than those presented in Refs. \cite{serbo_muonium,serbo_posi} for the ($e^+ e^-$) and ($\mu^+ \mu^-$) states. We believe that this difference is associated to the distinct treatments of the nuclear photon spectrum. 
Regarding to the event rates per year for heavy ion collisions, 
we predict that the number of   events  will be $\gtrsim 10^8/10^3$ for  the ($e^+ e^-$) / ($\mu^+ \mu^-$) production.
The predictions for the ($\tau^+ \tau^-$) states are presented here for the first time. It is important to emphasize that differently from the ($\mu^+ \mu^-$) states, which annihilate long before its constituents weakly decay, the ($\tau^+ \tau^-$) annihilation and the weak $\tau$ decay compete and, consequently, the tauonium cannot be considered a genuine QED bound state. Our results indicate that the associated cross sections are nine / three orders of magnitude smaller than for ($e^+ e^-$) / ($\mu^+ \mu^-$) states and that the number of events per year will be small for the paratauonium and negligible for the orthotauonium. These aspects become the experimental reconstruction  of the single and triplet ($\tau^+ \tau^-$) final states a hard task.

\begin{figure}[t]
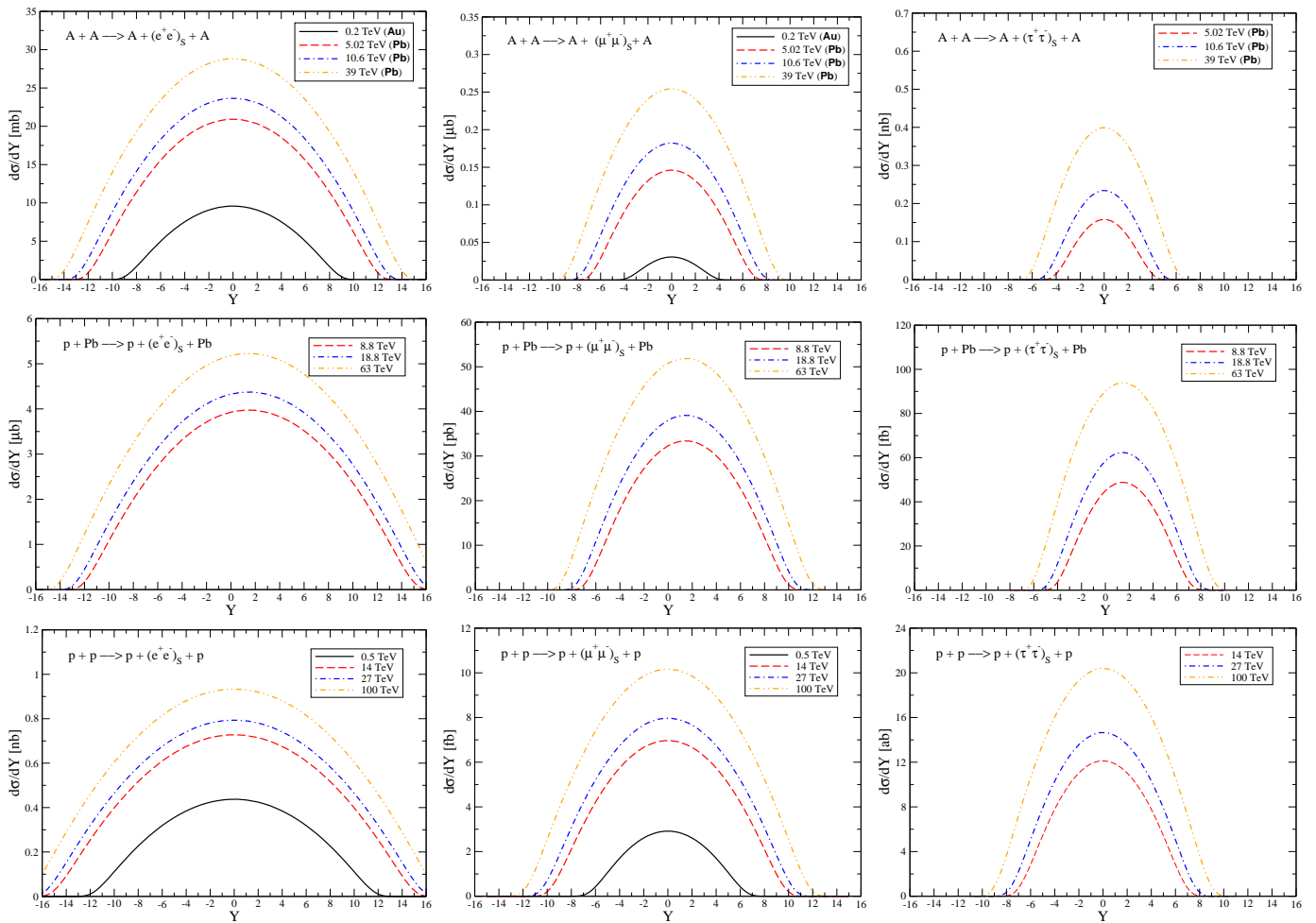

	\centering
	\begin{tabular}{ccc}
	\includegraphics[width=0.33\textwidth]{dist_Y_AA_positronium_born.eps} & \includegraphics[width=0.33\textwidth]{dist_Y_AA_muonium.eps} &
\includegraphics[width=0.33\textwidth]{dist_Y_AA_tauonium.eps} \\
		\includegraphics[width=0.33\textwidth]{dist_Y_pPb_positronium.eps} &  \includegraphics[width=0.33\textwidth]{dist_Y_pPb_muonium.eps} & 
		\includegraphics[width=0.33\textwidth]{dist_Y_pPb_tauonium.eps} \\
				\includegraphics[width=0.33\textwidth]{dist_Y_pp_positronium.eps} 
			 & \includegraphics[width=0.33\textwidth]{dist_Y_pp_muonium.eps}  & \includegraphics[width=0.33\textwidth]{dist_Y_pp_tauonium.eps} 
			\end{tabular}
	\caption{Rapidity distributions for the photoproduction of singlet QED bound states in AA (upper panels), pA (middle panels) and pp (lower panels)  collisions. }
	\label{fig:rapidity}
\end{figure}

The predictions presented in Table \ref{table:AA} indicate that the photoproduction of singlet QED bound states will be dominant. This result motivates a more detailed analysis of this final state, taking into account of the rapidity range usually covered by central and forward detectors. Considering that the photon energies  $\omega_i$ can be expressed in terms of  the rapidity $Y$ of the final state and $W_{\gamma \gamma}$ as follows: 
\begin{eqnarray}
\omega_1 = \frac{W_{\gamma \gamma}}{2} e^Y \,\,\,\,\mbox{and}\,\,\,\,\omega_2 = \frac{W_{\gamma \gamma}}{2} e^{-Y} \,\,\,,
\label{ome}
\end{eqnarray}
one has that the rapidity distribution for the photoproduction of a singlet QED bound state can be expressed by
\begin{eqnarray}
\frac{d\sigma_S}{dY}   
&=& \int \mbox{d}^{2} {\mathbf r_{1}}
\mbox{d}^{2} {\mathbf r_{2}} 
\mbox{d}W_{\gamma \gamma} 
 \frac{W_{\gamma \gamma}}{2} \, \hat{\sigma}\left(\gamma \gamma \rightarrow (l^+ l^-)_S ; 
W_{\gamma \gamma} \right )  N\left(\omega_{1},{\mathbf r_{1}}  \right )
 N\left(\omega_{2},{\mathbf r_{2}}  \right ) S^2_{abs}({\mathbf b})  
  \,\,\, .
\label{cross-sec-2}
\end{eqnarray}
Such expression allow us to estimate the rapidity distributions for the $(l^+ l^-)_S$ production in pp, pA and AA  collisions at RHIC, LHC, HE -- LHC and FCC. The resulting predictions are presented in Fig. \ref{fig:rapidity}. One has that the rapidity distributions are symmetric for pp and AA collisions. On the other hand, they are asymmetric for pA collisions due to the difference between the photon spectra for the proton and nucleus. Moreover, the distributions are wider in rapidity for the parapositronium in comparison to those for the $(\mu^+ \mu^-)_S$ and $(\tau^+ \tau^-)_S$ states. The predictions for the cross sections in the central ($ -2.5 \le Y \le 2.5$) and forward ($ 2.0 \le Y \le 4.5$) rapidity ranges are presented in Table \ref{table:cspara}. One has that the predictions for the parapositronium in pp/pA/AA collisions are of the order of  nb/$\mu$b/mb, respectively, with the results for forward rapidities being a factor $\approx$ 2 smaller than for the central rapidity range. For the paramuonium (paratauonium), the predictions are  five (nine) orders of magnitude smaller than for the $(e^+ e^-)_S$ production. Assuming that the expected integrated luminosity for pp collisions at RHIC, LHC, HE -- LHC and FCC is fb$^{-1}$, we predict that the number of events associated to the production of parapositronium will be $\approx 10^6$/year. In contrast, some dozens of $(\mu^+ \mu^-)_S$ states will be produced per year. For pA collisions, the cross sections are enhanced by a factor $Z^2$, but the integrated luminosity is expected to be 1.0 (29.0) pb$^{-1}$ in pPb collisions at the LHC (FCC). As a consequence,  the number of events per year associated to the $(e^+ e^-)_S$ [$(\mu^+ \mu^-)_S$] production will be larger than $10^7\, [10^2]$. Finally, for AA collisions, we predict that $\approx 10^9/\,10^3$ $(e^+ e^-)_S$ / $(\mu^+ \mu^-)_S$ states will be produced per year. For the paratauonium, we predict a negligible number of events in future pp, pA and AA collisions.

\begin{center}
	\begin{table}[t]
		\resizebox{17cm}{!}{
		\begin{tabular}{|c|c|c|c|c|c|c|}
			\hline
			\hline 
			& \multicolumn{2}{c|}{\bf Parapositronium}          &        \multicolumn{2}{c|}{\bf Paramuonium} &        \multicolumn{2}{c|}{\bf Paratauonium}   \\
			\hline 	
			& $ -2.5 \le Y \le 2.5$ & $ 2.0 \le Y \le 4.5$ & $ -2.5 \le Y \le 2.5$ & $ 2.0 \le Y \le 4.5$ & $ -2.5 \le Y \le 2.5$ & $ 2.0 \le Y \le 4.5$ \tabularnewline
			\hline 
pp ($\sqrt{s_{NN}} = 0.5$ TeV)  & 2.2$\times 10^{6}$ & 1.0$\times 10^{6}$ & 13.7  & 5.2 &  11.9 $\times 10^{-3}$   & 2.0 $\times 10^{-3}$   \tabularnewline
			\hline 
pp ($\sqrt{s_{NN}} = 14$ TeV)& 3.6$\times 10^{6}$ & 1.7$\times 10^{6}$ & 34.0  & 15.3 &  57.7 $\times 10^{-3}$  & 22.8 $\times 10^{-3}$  \tabularnewline
			\hline 
pp ($\sqrt{s_{NN}} = 27$ TeV) & 3.9$\times 10^{6}$ & 1.9$\times 10^{6}$ & 39.0  & 17.8 &  70.4 $\times 10^{-3}$  & 29.1 $\times 10^{-3}$    \tabularnewline
\hline
pp ($\sqrt{s_{NN}} = 100$ TeV) & 4.6$\times 10^{6}$ & 2.2$\times 10^{6}$ & 50.0  & 23.3 &  99.1 $\times 10^{-3}$   & 43.5 $\times 10^{-3}$  \tabularnewline
\hline 
\hline
\hline
pPb ($\sqrt{s_{NN}} = 8.8$ TeV)  & 19.4$\times 10^{9}$ & 9.7$\times 10^{9}$ & 157.0$\times 10^{3}$  & 78.6$\times 10^{3}$ &  206.3   & 105.6   \tabularnewline
			\hline 
pPb ($\sqrt{s_{NN}} = 18.8$ TeV)& 21.4$\times 10^{9}$ & 10.7$\times 10^{9}$ & 184.7$\times 10^{3}$  & 92.9$\times 10^{3}$ &  273.4   & 139.0   \tabularnewline
			\hline 
pPb ($\sqrt{s_{NN}} = 63$ TeV) & 25.7$\times 10^{9}$ & 12.9$\times 10^{9}$ & 248.4$\times 10^{3}$  & 124.8$\times 10^{3}$ &  431.1   & 217.8   \tabularnewline
			\hline 
			\hline 
			\hline 
AuAu ($\sqrt{s_{NN}} = 0.2$ TeV)  & 46.6$ \times 10^{12}$ & 20.4$ \times 10^{12}$ & 125.5 $\times 10^{6}$  & 19.7 $\times 10^{6}$ &  3.8 $\times 10^{3}$  & 3.6   \tabularnewline
			\hline 
PbPb ($\sqrt{s_{NN}} = 5.02$ TeV)& 103.1$ \times 10^{12}$ & 48.2$ \times 10^{12}$ & 693.9 $\times 10^{6}$  & 269.3 $\times 10^{6}$ &  671.0 $\times 10^{3}$   & 130.2 $\times 10^{3}$  \tabularnewline
			\hline 
PbPb ($\sqrt{s_{NN}} = 10.6$ TeV) & 116.7$\times 10^{12}$ & 55.0$ \times 10^{12}$ & 874.5 $\times 10^{6}$ & 359.4 $\times 10^{6}$ &  1044.0 $\times 10^{3}$ & 280.0 $\times 10^{3}$   \tabularnewline
\hline
PbPb ($\sqrt{s_{NN}} = 39.0$ TeV) & 142.5$ \times 10^{12}$ & 67.9$ \times 10^{12}$ & 1236.0 $\times 10^{6}$  & 539.9 $\times 10^{6}$ &  1867.0 $\times 10^{3}$   & 663.1 $\times 10^{3}$   \tabularnewline			
			\hline 			
			\hline 
		\end{tabular}
		}
		\caption{Cross sections (in fb)  for the photoproduction of singlet QED bound states in pp, pA and AA collisions considering the central and forward rapidity ranges.}
		\label{table:cspara}
	\end{table}
\end{center}

\begin{figure}[t]
	\centering
	\begin{tabular}{ccc}
	\includegraphics[width=0.45\textwidth]{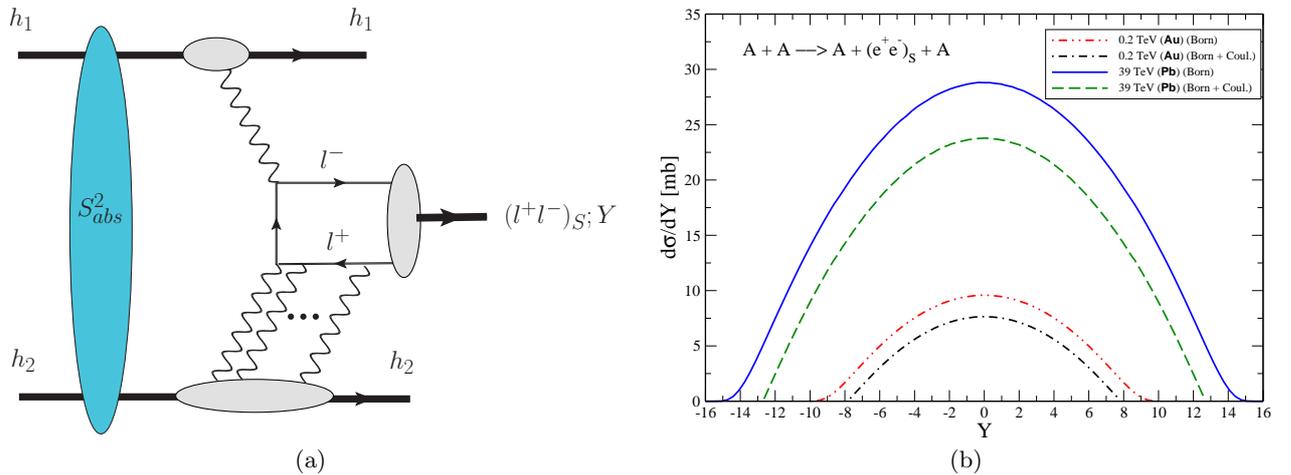} & \,\,\,\,\, & \includegraphics[width=0.45\textwidth]{born_X_CC.eps} \\
	(a) & \, &  (b) 
			\end{tabular}
\caption{(a) Coulomb corrections for the photoproduction of singlet QED bound states. (b) Rapidity distributions for the photoproduction of a parapositronium in heavy ion collisions calculated at the Born level and including the Coulomb corrections.}
\label{fig:diagramCC}
\end{figure}

The previous results were derived considering the cross sections at the Born level, disregarding higher order corrections in the parameter $\nu = Z \alpha$, which becomes of the order of 1 for heavy nuclei. {  In the last decades, the calculation of these corrections for the QED pair production has been performed by several groups and the importance of its contribution for the total cross section was a theme of intense debate (See, e.g., Refs. 
\cite{Segev:1997yz,Baltz:1998zb,Eichmann:1998eh,Lee:1999eya,Baltz:2001dp,Lee:2001ea,Bartos:2001jz,Aste:2001te,Baltz:2003dy,Baltz:2004dz,Hencken:2006ir,Baltz:2007gs,Baltz:2009fs}). As reviewed in Ref. \cite{upc_dilepton}, these corrections are analogous to the Bethe - Heitler process on a heavy target and can be estimated using the approach developed in Refs. \cite{Ivanov:1998ru,Ivanov:1998ka}. Recent results \cite{Zha:2021jhf}  indicate that the current data for the QED pair production in ultraperipheral heavy ion collisions are quite well described if these higher order corrections are taken into account. For the photoproduction of QED bound states, the higher order corrections were estimated  in Refs. \cite{serbo_pra,serbo_muonium,serbo_posi}, which extended  the formalism presented in Refs. \cite{Ivanov:1998ru,Ivanov:1998ka} for this final state. In what follows, we will consider this approach in our calculations. In particular, }  we will estimate the impact of the Coulomb correction (CC), associated to the multiphoton exchange of the produced system with nuclei, on our predictions for the photoproduction of a singlet QED bound state in heavy ion collisions. A typical diagram considered in our analysis is represented in Fig. \ref{fig:diagramCC} (a). As demonstrated in Refs. \cite{serbo_pra,serbo_muonium,serbo_posi}, the Coulomb corrections decrease the cross sections, with its magnitude  being  suppressed with the increasing of the center - of - mass energy and for larger masses of the QED bound state. In particular, the results presented in Ref. \cite{serbo_muonium} indicate that these corrections already become negligible for the $(\mu^+ \mu^-)_S$ production. Consequently, we will focus on the parapositronium production. One has estimated the Coulomb corrections using the formalism developed in Refs. \cite{serbo_pra,serbo_posi} and included its contribution in the calculation of the associated rapidity distribution. Our results are presented in Fig.  \ref{fig:diagramCC} (b)  considering heavy ion collisions at the RHIC and FCC energies. The predictions derived at the Born level are also presented for comparison. As expected, the Coulomb corrections reduce the rapidity distributions. The predictions for the cross sections and event rates per year considering the rapidity ranges covered by the central and forward detectors are presented in Table \ref{tab:cc}. In comparison to the results presented in Table \ref{table:cspara}, the predictions for the RHIC energy are reduced by $\approx 22 \%$, while  the suppression for the FCC is  $\approx 18 \%$. Although the number of events is also reduced, it still is very large ($\gtrsim 10^8$). Our results indicate that a future experimental  analysis of the parapositronium production in heavy ion collisions can be useful to constrain the treatment of the Coulomb corrections.

\begin{table} [t]
	\centering
	\begin{tabular}{|c||c|c||}
	\hline
	 & $-2.5 \le Y \le 2.5$   & $2.0 \le Y \le 4.5$   \\ \hline \hline
	AuAu\;($\sqrt{s_{NN}}$\;=\;0.2\;TeV)  & 36.8 $\times 10^{12}$ (3.6$\times 10^{8}$)  & 15.6 $\times 10^{12}$ (1.6$\times 10^{8}$)\\ \hline
	PbPb\;($\sqrt{s_{NN}}$\;=\;5.02\;TeV) & 83.1 $\times 10^{12}$ (8.3$\times 10^{8}$)  & 38.2 $\times 10^{12}$ (3.8$\times 10^{8}$)\\ \hline
	PbPb\;($\sqrt{s_{NN}}$\;=\;10.6\;TeV) & 94.9 $\times 10^{12}$  (9.5$\times 10^{8}$)  & 44.0 $\times 10^{12}$ (4.4$\times 10^{8}$)\\ \hline
	PbPb\;($\sqrt{s_{NN}}$\;=\;39\;TeV)   & 117.3 $\times 10^{12}$   (13.0$\times 10^{9}$)& 55.3 $\times 10^{12}$ (6.1$\times 10^{9}$)\\ \hline
	
	\end{tabular}
	\caption{Cross sections in fb (Event rates per year) for the photoproduction of a parapositronium estimated including the Coulomb corrections.}
	\label{tab:cc}
	\end{table}

{ The results presented above indicate  that the associated  number of events for the current and future colliders will be large, in particular for the production of the parapositronium and paramuonium states. Such large values strongly motivate the discussion of two important aspects: (a) the importance of the backgrounds, and (b) how feasible is the experimental separation of the events. Regarding the first aspect, one has that the dominant decay channel of the singlet states is the decay into two - photons with a small invariant mass. As a consequence, the more important background will be the diphoton system generated in the light -- by -- light (LbL) scattering. Assuming that the associated LbL cross section is known and  constrained by the recent data, such background could be removed, allowing access to the events  associated to the production of singlet QED bound states. On the other hand, the second aspect is a challenge, especially for parapositronium states. For this final state,  the two-photons produced in its decay have a very small invariant mass ($m_{\gamma \gamma}$), which is out of the reach of the current colliders and it is not clear if this kinematical range could be accessed in a near future. Events with $m_{\gamma \gamma} \gtrsim 100$ MeV are expected to be reconstructed by the ALICE - FoCal \cite{ALICE:2020mso} and LHCb \cite{LHCb:2018roe} detectors, which implies that, in principle, events associated to the production of paramuonium states could be studied. However, in order to derive  predictions for the kinematical range covered by these detectors, it is fundamental to include the treatment of this final state in the current Monte Carlo codes dedicated to ultraperipheral heavy ion collisions as. e.g. the StarLight and SuperChic event generators \cite{starlight,super}. In addition, it is important to improve the description of the intrinsic transverse  momentum of the incoming photons (See e.g. Refs. \cite{Zha:2018tlq,Brandenburg:2021lnj}), which  affect the $p_T$ distributions of the decay products.  Both aspects deserve a more detailed analysis which we intend to perform in the near future.    }

\section{Summary}

In this  study we have investigated the photoproduction of  singlet and triplet QED bound states in pp, pA and AA collisions at different center -- of -- mass energies. Our main motivation was to estimate the associated cross sections and rapidity distributions in order to verify if these final states could be measured in the future, which would allow us to improve our understanding about the positronium and observe, for the first time, the muonium and tauonium states. Our results demonstrated that the singlet and triplet cross sections increase with the center - of - mass energy, with the production of singlet states being dominant. Moreover, we have presented predictions for the singlet cross sections considering the rapidity ranges covered by central and forward detectors, which indicate that the measurement of the paratauonium will be a hard task in the current and future colliders. In contrast, the cross sections and event rates for the production of the parapositronium and paramuonium states are large. Moreover,  we have studied the impact of the Coulomb corrections on the cross sections for the production of the parapositronium in heavy ion collisions and demonstrated that they are not negligible and that future data can be useful to constrain the treatment of these corrections. The results obtained in this paper strongly motivate the generalization of the current Monte Carlo generators used to simulate the ultraperipheral hadronic collisions \cite{starlight,super} in order to simulate the photoproduction of QED bound states, which would allow us to include the experimental cuts usually considered by the experimental collaborations. Such next step is currently in development and we intend to present the results in a forthcoming study.

\begin{acknowledgments}
This work was  partially financed by the Brazilian funding
agencies CNPq, CAPES,  FAPERGS, FAPESC and INCT-FNA (process number 
464898/2014-5).
\end{acknowledgments}

\hspace{1.0cm}

\end{document}